\documentclass[journal=jacsat,manuscript=article,layout=twocolumn]{achemso}

\newcommand*{\sometext}{We investigate a nanoscale dielectric capacitor model consisting of two-dimensional, hexagonal h-BN layers placed between two commensurate and metallic graphene layers using self-consistent field density functional theory. The separation of equal amounts of electric charge of different sign in different graphene layers is achieved by applying electric field perpendicular to the layers. The stored charge, energy, and the electric potential difference generated between the metallic layers are calculated from the first-principles for the relaxed structures. Predicted high-capacitance values exhibit the characteristics of supercapacitors. The capacitive behavior of the present nanoscale model is compared with that of the classical Helmholtz model, which reveals crucial quantum size effects at small separations, which in turn recede as the separation between metallic planes increases.

{\bf Keywords:} nanoscale dielectric capacitor, supercapacitor, energy storage, nanocapacitor, graphene, boron nitride}

\let\oldmaketitle\maketitle
\let\maketitle\relax

\author{V. Ongun \" Oz\c celik}
\affiliation{UNAM-National Nanotechnology Research Center, Bilkent University, 06800 Ankara, Turkey}
\alsoaffiliation{Institute of Materials Science and Nanotechnology, Bilkent University, Ankara 06800, Turkey}
\email{ongunozcelik@bilkent.edu.tr}
\author{S. Ciraci}
\affiliation{UNAM-National Nanotechnology Research Center, Bilkent University, 06800 Ankara, Turkey}
\alsoaffiliation{Institute of Materials Science and Nanotechnology, Bilkent University, Ankara 06800, Turkey}
\alsoaffiliation{Department of Physics, Bilkent University, Ankara 06800, Turkey}
\email{ciraci@fen.bilkent.edu.tr}

\title{Nanoscale Dielectric Capacitors Composed of Graphene and Boron Nitride Layers: A First Principles Study of High-Capacitance at Nanoscale}

\begin{document}

\twocolumn[
\begin{@twocolumnfalse}
\oldmaketitle
\begin{abstract}
\sometext
\end{abstract}
\end{@twocolumnfalse}
]

\section{Introduction}

Conventional (electrostatic) and electrochemical (EC and/or electrochemical double layer, EDLC) capacitors, batteries and fuel cells are systems which are widely used for energy storage.\cite{winter2004,pandolfo2006} As far as the combination of specific power versus specific energy capability is concerned, electrostatic capacitors keep an important place among them. On the other side, conventional capacitors are limited by the device geometry and the batteries are limited by the slow response time due to the motion of the ions in electrochemical reactions.

Recent developments in nanoscale physics have further widened our perception of energy storage mechanisms in materials. In this respect, recyclable, efficient and high capacity energy storage through light weight nanoscale mediums have attracted interest.\cite{salim1,salim2} In fact, capacitors at nanoscale have been developed as one of the most promising energy storage mediums, whereby several orders of magnitude higher energy densities have been realized. They are able to store and release charge faster, can deliver higher amounts of charge at higher power rates as compared to conventional batteries and have longer life with short load cycles which make them advantageous in various applications.\cite{winter2004,pandolfo2006,stoller2008,burke2000,holme2012,zeng2009,spaldin2006} Several materials such as mixed metal oxides,\cite{toupin2004} polymers,\cite{rudge1994} and carbon nanotubes\cite{liu2008, niu1997} have been used to fabricate  supercapacitors. Concomitantly, the performances of the latter were further improved and stabilized by using nonaqueous solvents with uniform translational diffusion coefficient.\cite{futaba2006} This important behavior was also observed in simulations of acetonitrile confined in carbon nanotubes, which makes it a suitable solvent.\cite{kalugin2008}

Recently, nanoscale dielectric capacitors (NDC) have rapidly developed and achieved properties, which are superior to other systems of energy storage. An ideal NDC should be composed of metallic layers that can store and release charge and a dielectric material in between these layers for increasing the capacitance value without increasing the dimensions of the structure. Recent theoretical and experimental studies on graphene, carbon nanotubes and metallic nanowires have focused on understanding the dielectric properties of these structures and forming thin layers that can serve as ideal charge holding metallic plates.\cite{spaldin2006, sorel2012, uprety2007, lee2008} A theoretical method has been developed for calculating the dielectric response of periodic metal-insulator heterostructures and hence the microscopic properties of thin-film capacitors modeled by Ag/MgO/Ag films.\cite{spaldin2006} In a recent study, nanostructured electrodes were used to fabricate transparent capacitors with polymer dielectrics and carbon nanotube electrodes\cite{sorel2012} which are important in a range of applications from sensors to transparent circuits. Similarly structural, optical and electrical properties of transparent carbon nanotube based capacitors on glass substrates were also examined and shown to be highly efficient in photovoltaic and solar energy storage devices.\cite{uprety2007} It was also experimentally demonstrated that Ag nanowires deposited on glass substrates can be used as electrodes\cite{lee2008} which can possibly be used in nanocapacitor fabrication.

Improving the permittivity of the dielectrics used have been also important for higher capacitance values of NDCs. With this regard, recent progress was made in increasing the permittivity and breakdown strength of nanocomposite materials by using fillers\cite{tang2013, kim2009} and high aspect ratio nanowires.\cite{tang2011} It was demonstrated that polyvinylidene fluoride and Ba-Ti-O  nanowires can both increase the breakdown strength of nanocomposite dielectrics and increase their energy densities.\cite{tang2013} The dependence of the energy density on the volume fraction was also investigated and maximum energy density values were achieved at 50\% nanoparticle volume fraction.\cite{kim2009} Furthermore, the existence of dielectric dead layer at nanoscales were investigated on $SrRuO_3/SrTiO_3$ dielectric capacitors and the reason behind the reduction in the capacitance values in experiments as compared to theoretical predictions were explained.\cite{spaldin2006}

Exceptional properties of graphene have been exploited actively for future nanoscale electronics and spintronic applications.\cite{geim2004,netto2009} Incidentally, it was shown that graphene can sustain current densities six order of magnitude larger than copper.\cite{geim2009} Graphene with its 2D one-atom thick honeycomb structure showing a perfect electron-hole symmetry and high chemical stability\cite{geim2007} has also been proposed as an ideal nanocapacitor material.\cite{yongchao2008, stoller2008, vivekchand2008} In previous studies, capacitance per unit area values of 80 $\mu$F/cm$^2$ and 394 $\mu$F/cm$^2$ have been achieved for capacitors with electrodes comprised of pristine graphene and multilayer reduced graphene oxide.\cite{yoo2011} Other than obtaining high capacitance values, graphene capacitors have also been a field of study for observing interaction phenomena in graphene layers.\cite{pnas2013}

\begin{figure}
\includegraphics[width=8cm]{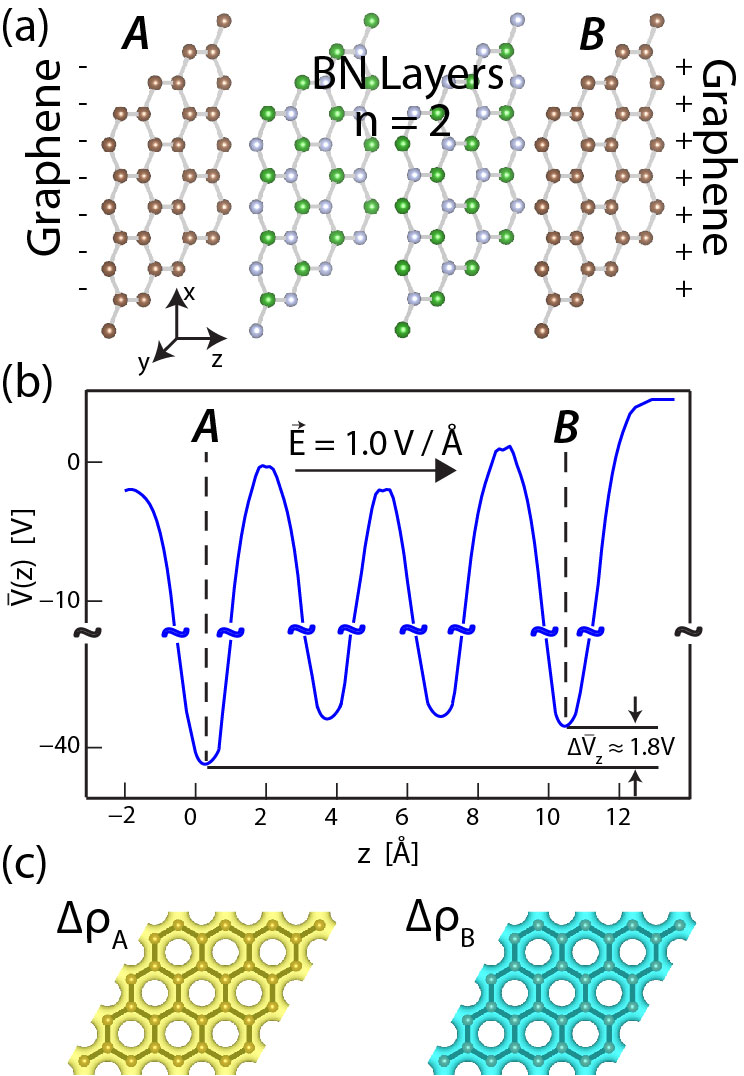}
\caption{(Color online) (a) A NDC model with $n=2$ h-BN layers serving as dielectrics are capped by two  parallel graphene layers serving as metallic plates. The whole system is subjected to a uniform electric field along the $z$-axis so that graphene plates are charged by surface charge densities of $ -\sigma $ and $+ \sigma $, respectively. (b) Schematic description of the calculated $(x,y)$-plane averaged electronic potential, $\bar{V}(z)$. The difference of the potential energy between graphene layers A and B is $e \bar{V}_{z}$. (c) Isosurfaces of the self-consistent difference charge densities of the negatively charged (A) and positively charged (B) graphene layers are $\Delta \rho_A$ and $\Delta \rho_B$, respectively. The isosurface values are taken as 0.01 electrons / \AA $^2$. Yellow and blue isosurfaces indicate excess and depleted electrons, respectively. The computations are performed on a $2 \times2 $ supercell with a vacuum spacing of 20 \AA.}
\label{fig1}
\end{figure}

\section{Model}
In this paper, we consider a NDC model envisaged from composite materials and study its capacitive behavior using \textit{ab initio} calculations within the density functional theory (DFT). Our nanoscale capacitor model is composed of hexagonal h-BN layers, which are stacked between two metallic graphene sheets as described in ~\ref{fig1} (a) and attains high gravimetric capacitance values. We compare the capacitance values obtained from the present first principles total energy calculations with those estimated within the classical Helmholtz model and reveal crucial quantum size effects at nanoscale.

Single, as well as multilayer h-BN are wide band gap insulators and they can serve as a dielectric material between metallic graphene layers. Additionally, sheets of h-BN multilayers that are lattice matched to graphene allow one to attain robust and high precision nanoscale spacings  between two parallel metallic graphene sheets, which can be set to desired values. It has already been shown both experimentally\cite{liu11} and theoretically\cite{ozcelik, katsnelson} that h-BN layers of any thickness can be grown on graphene layers and vice versa. Additionally, using phonon dispersion calculations and molecular dynamics simulations, stabilities of such structures and other similar carbon and BN allotropes, namely graphyne and BNyne, were recently studied.\cite{topsakal2009, graphyne} Other than bilayer structures, it is also possible to grow perpendicular carbon and BN chains on top of graphene and single layer h-BN.\cite{can_chain, ongun_chain}

Controlling the spacing between metallic plates at nanoscale is the crucial feature in attaining high capacitances. Series, parallel, mixed and 3D combinations of these structures, which can also be fabricated by repeated stacking of varying numbers of graphene and h-BN layers, offer number of options in constructing novel NDCs with diverse functions. Thus, by varying the separation distance in our capacitor model, we observe interesting quantum size effects at small separations which recede as the distance between the graphene plates increases. Since the principle objective of using a capacitor is to store energy by storing equal magnitude of electric charges of opposite sign in two disconnected conducting plates, a charged capacitor is in a static and non-equilibrium state. The energy stored this way is released when the plates are connected to a circuit, so that the discharged capacitor turns into an equilibrium state. At nanoscales, since the separation thickness of the devices can be as small as a nanometer, the stored energy has to be calculated from the first-principles. However, the available first-principles methods allow us to treat the distribution of only one kind of excess charge (positive or negative) in the same system at a time.\cite{chan2011, chan20112, suarez2011, topsakal2012} Therefore, the main obstacle in the present study was separating positive and negative charges on the plates of a capacitor. In our model, the charge carriers are electrons themselves; while they exist in excess in one plate, they are depleted from the other one. Thus, our paper treats the charged capacitor. The charge separation is achieved by applying an external electric field $\vec{E}$,  perpendicular to the graphene layers, as schematically shown in ~\ref{fig1} (a). This situation mimics the operation of a capacitor, whereby the surface charge of opposite sign initially stored on different metallic plates create a perpendicular electric field. The shorting of these two plates and hence the discharge of the capacitor are hindered by placing sufficient amount of vacuum space or dielectric h-BN layers between graphene plates.

On the other side, charge separation through applied electric field by itself brings along a serious problem: When treated using periodic boundary conditions (PBC), the vacuum potentials between the periodically repeating negatively charged surfaces or systems under an electric field form a quantum well. If this quantum well dips below the Fermi level, electrons start to be accommodated in this quantum well in calculations using plane wave (PW) basis set. Accordingly, excess electrons in the graphene layers become vulnerable to spurious charge spilling into the vacuum space between supercells. Therefore, this artifact sets a limit to the amount of negative charging that can be treated by PW for a given width of the vacuum region $s$. Nevertheless, this artifact of PW methods can be circumvented by using a local basis set such as atomic orbitals (AO), since they fail to represent the states which can be bound to the quantum well in the middle of the vacuum spacing. In recent studies\cite{vacuum, topsakal2013} the treatment of charged systems and systems under electric field has been extensively analyzed. Thus, treating an NDC from the first-principles using periodic boundary conditions, where electric charges of opposite sign are separated through an applied perpendicular electric field, is the unique aspect of our study.

\begin{figure*}
\includegraphics[width=16cm]{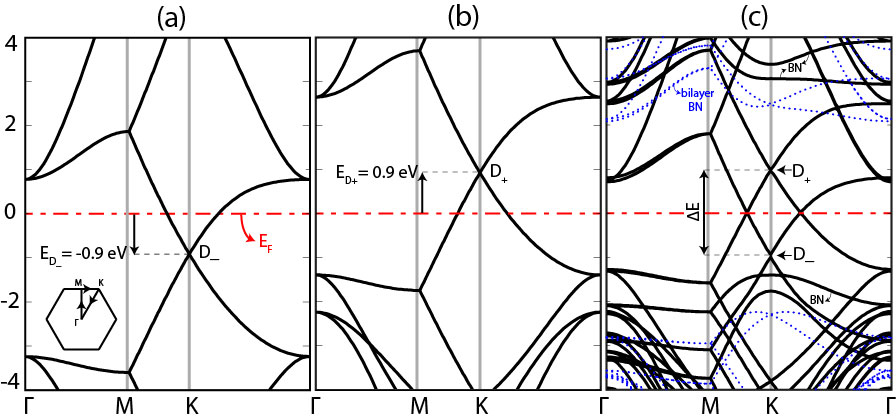}
\caption{(Color online) Electronic energy band structures calculated for ($2 \times 2$) supercell. The zero of energy is set to the Fermi level $E_F$ shown by the red dash-dotted lines. (a) Isolated, single layer graphene, which is negatively charged  by $Q=-0.06e$ per primitive unitcell (or $\sigma = -0.18 C/m^2$). (b) Positively charged single layer graphene with $Q=+0.06e$ per primitive unitcell (or $\sigma = +0.18 C/m^2$). $E_{D_\pm} \simeq \pm $ 0.9 eV are the down and up shifts of the Dirac points ($D_{-}$ and $D_{+}$) relative to the Fermi level for negative and positive charging, respectively. (c) The electronic band structure of the capacitor consisting of two h-BN layers capped by single layer graphenes, which is subjected to an electric field of $\vec{E} =1.0$ V/\AA. The band structure of the free h-BN bilayer is shown by the blue dotted lines.  The energy difference between $D_{-}$ and $D_{+}$ points is $\Delta E$. The inset shows the symmetry directions in the central parallel plane including $\Gamma$-point.}

\label{fig2}
\end{figure*}

\section{Method}
In view of the above discussions, we carried out first-principles spin-polarized and spin-unpolarized calculations within density functional theory (DFT) using atomic orbitals (AO) as local basis set. The exchange-correlation potential is approximated by the generalized gradient approximation (GGA) using Perdew, Burke and Ernzerhof (PBE) functional.\cite{pbe}. The eigenstates of the Kohn-Sham Hamiltonian are expressed as linear combinations of numerical atomic orbitals. A 200 Ryd mesh cut-off is chosen and the self-consistent field (SCF) calculations are performed with a mixing rate of 0.1. Core electrons are replaced by norm-conserving, nonlocal Troullier-Martins pseudopotentials.\cite{troullier1991} The convergence criterion for the density matrix is taken as 10$^{-4}$. Atomic positions, lattice constants are optimized using the conjugate gradient method, where the total energy and atomic forces are minimized. In particular, the minimum energy stacking sequence of the composite material consisting of $n$ BN layers between two graphene is determined for each $n$. All numerical calculations are performed using the SIESTA code.\cite{siesta} Dipole corrections\cite{payne} are applied in order to remove spurious dipole interactions between periodic images for the neutral calculations. Calculations are carried out on the ($2 \times 2$) supercells in order to account possible reconstructions, but the relevant values are given per primitive cell. The dielectric constant values of stacked boron nitride layers were calculated using plane wave methods as implemented in the PWSCF package.\cite{pwscf}

Our previous analysis\cite{mos2} have shown that the interlayer spacing of graphite has been overestimated by GGA approximation using PW91 functional\cite{pw91}, but has improved to near 4\% of the experimental value with the van der Waals (vdW) correction.\cite{vdw} To reveal the accuracy of the spacing between graphene and single layer h-BN; and also the spacing between the graphene layers capping two h-BN layers in ~\ref{fig1}(a) we optimize the structures with two different methods: (i) with PBE using SIESTA code and (i) with PBE including vdW correction using VASP code.\cite{vasp} While the spacing between single graphene and h-BN layers is overestimated by 5.5\% in PBE relative to that of obtained by including vdW correction, this overestimation of the spacing between the graphene layers in our capacitor model reduced to 1.5\%. As for the lattice constant of the hexagonal primitive unit cell predicted by PBE (SIESTA) and PBE+vdW (VASP), the former method overestimates the lattice constant of the free graphene (free h-BN) by only 0.6\% (0.7\%). In view of the fact that h-BN layers serve as dielectrics, the overestimation of the interlayer spacings by approximately 5.5\% is acceptable.

Regarding the structure, free h-BN and free graphene have lattice constants calculated to be $a$= 2.53\AA~ and $a$=2.48\AA~, respectively, which lead to a lattice mismatch of $\sim$2\%. When grown on top of graphene, this lattice mismatch is accommodated by a supercell of 50$\times$50 comprising 10.000 atoms. Therefore, the investigation of superlattice effects and miniband formation may not be investigated easily from first-principles calculations. However, the effects of small lattice mismatch is expected to be minute\cite{wallbank} to affect the properties of dielectric h-BN spacer. Additionally, when optimized together in a single unit cell, h-BN having relatively smaller in-plane stiffness\cite{ansiklopedi} is compressed to the smaller unit cell.

\section{Results and discussion}

\subsection{Electronic Structures}

We first consider two isolated graphene layers where each one is charged by $\pm Q=$ 0.06 electrons per primitive unitcell (or $\pm \sigma = 0.18 C/m^2$). The electronic band structure of each graphene is shown in ~\ref{fig2} (a)-(b): The Dirac point $D_{-}$ dips below the Fermi level when a the isolated graphene is negatively charged by $\sigma=-0.18 C/m^2$. In contrast, the Dirac point $D_{+}$ raises above the Fermi level when the isolated graphene layer is positively charged or hole doped by $\sigma= 0.18 C/m^2$. The upward and downward shifts of the corresponding band structures in ~\ref{fig2} (a)-(b) are equal and $|E_{D_{-}}| = E_{D_{+}} \simeq$ 0.9 eV.

Next, we examine our model of nanocapacitor described in ~\ref{fig1}(a), which is actually one single system of a composite material consisting of two insulating h-BN layers placed between two parallel semimetallic graphene layers. Initially the graphene layers have zero net charge. However, when exerted by a positive $\vec{E}$ along the $z$-axis, self-consistent field calculations accommodate excess electrons on the left graphene while the same amount of electrons are depleted from the right graphene. Hence, the integral of the volume charge in the cell of the capacitor is still zero; $\int_{cell}QdV=0$. The atomic structures, interlayer spacings, relative positions of the layers and the cell parameters, are optimized. The plane-averaged electronic potential, $\bar{V}(z)$, is shown in ~\ref{fig1}(b) under an external electric field of $\vec{E} =$ 1.0 V/\AA. For this case, the potential energy difference between the two graphene layers is calculated to be $ e \Delta \bar{V}_z \simeq$ 1.8 eV, which leads to the accumulation of equal amount of surface charge of opposite sign, $\pm Q=$ 0.06 electrons per primitive unitcell (or $\pm \sigma = 0.18 C/m^2$) on either graphene layers. The isosurface plots of the difference charge density of the negatively charged (A) and positively charged (B) graphene layers, namely the difference of charge densities between the charged ($\rho_{A,B}$) and neutral ($\rho_{0}$) graphene layers, $\Delta \rho_{A,B} = \rho_{A,B} - \rho_{0}$ illustrate the charge separation.

The equilibrium band structure of this optimized capacitor in a ($k_{x},k_{y},k_{z}=0$) symmetry plane including the center of the Brillouin zone parallel to the graphene and BN planes of capacitor is presented in~\ref{fig2}(c). Owing to the weak coupling between graphene layers and BN-layers, the band structure of the capacitor is practically the combination of the bands in~\ref{fig2} (a) and (b), and those of h-BN bilayer as shown by the dotted lines in ~\ref{fig2}(c), except that the insulating bands of each h-BN layers shift relative to each other. Based on the calculated electronic structure we make following comments:

(i) Being an insulator and serving as a spacer and dielectric, the variation of the band structure and band gap of BN with applied electric field $\vec{E}$ is closely related to the breakdown voltage (i.e. the maximum voltage occurring between the plates before they are shorted internally) and dielectric strength (i.e. the maximum electric field magnitude that an insulator can withstand without the occurrence of breakdown). The band gap of free single layer h-BN, which was calculated to be $\sim$4.6eV,\cite{topsakal2009} rises up to 6.8 eV after GW self-energy correction.\cite{ansiklopedi}. Experimentally, it is measured to be $\sim5.5$eV.\cite{watanabe2004} Earlier it was demonstrated that the band gap of a single layer MoS$_2$ is overestimated by GW correction.\cite{gwmos2,mx2}

(ii) The dielectric spacer undergoes a potential difference between two charged metallic plates, whereby the band gap is bent. For a given uniform $\vec{E}$ exerted perpendicular to the plates, the band bending is approximately proportional with the width of dielectric spacer and the magnitude of the electric field. Upon bending of the bands in direct space, the band gap between the conduction and the valance band decreases with increasing potential difference. In the \textbf{k}-space the band bending manifests itself as the energy shift of the bands of each parallel h-BN layer. For example, the energy shifts of consecutive bands is calculated to be $\sim$ 1 eV at $\vec{E}$=1 V/\AA~ for 5 layers of BN without the capping graphene layers. Even if the band gap in each layer is practically unaltered, the bad gap of the multilayered h-BN decreases with increasing $\vec{E}$ since the bottom of the conduction band is lowered along one direction while top of the valance band raises along the opposite direction to narrow down the band gap. At the breakdown voltage of the BN spacer comprising finite number of h-BN, the Fermi level overlaps the band of capping graphene (or the top of the valance band of the bare BN spacer in the absence of capping graphenes) with the lowest conduction band, which is bent to lower energies.  Thereafter, a resonant tunneling sets in between negatively and positively charged metal plates through the h-BN layers. This physical event has been identified as the Zener breakdown,\cite{zener} whereby a wide band gap insulator becomes a conductor. The tunneling current exponentially decays with the increasing hight and width of tunneling barrier. In this respect, the nanoscale dielectric capacitor proposed here having small spacings between the plates allows a high surface charge density $\sigma$ on the metallic plates and hence induces large perpendicular electric field in the spacer. The energy barrier, $\Delta \Phi$, between the Fermi level and the lowest conduction band of BN spacer, is the tunneling barrier. Higher $\Delta \Phi$, by itself, hinders the tunneling of carriers from one plate to other through the BN spacer.

(iii) At the nanoscale, the dielectric strength can be deduced from the band structure of BN spacer, which is stronger than the dielectric strength of BN crystal. For the nanoscale graphene+BN, planar capacitor like the one presented in ~\ref{fig1}, the breakdown voltage is estimated from the band gap of a single layer h-BN which is $\sim$ 3 V for $n=3$. We note, however, that for BN spacer consisting of a few layers, the breakdown voltage depends on various conditions, such as non-equilibrium state, the coupling between graphene and adjacent h-BN, etc and requires a thorough analysis.

(iv) The integrals of the SCF charge on the graphene layers A and B in ~\ref{fig2}(c) are equal to the excess charge values used in the calculation of ~\ref{fig2}(a) and (b), namely $\int^{B,A} \sigma(x,y) dx dy = \pm Q=$ 0.06 electrons per primitive cell (or $\pm \sigma = 0.18$ $C/m^2$). (iii) The calculated energies of the Dirac points for the left graphene layer, ($E_{D_{-}}$) in ~\ref{fig2}(a) and the right graphene layer, $E_{D_{+}}$ in ~\ref{fig2}(b) relative to the common Fermi level $E_F$ are equal in magnitude because of the electron-hole symmetry near the Dirac points, i.e. $|E_{D_{-}}| = E_{D_{+}}$.  In the electronic band structure of the capacitor in ~\ref{fig2}(c), the energy difference between the Dirac points $D_{+}$ and $D_{-}$ is $\Delta E$. Interestingly, $\Delta E$ is equal the sum of the energy shifts in ~\ref{fig2}(a) and (b), i.e. $\Delta E = |E_{D_{-}}| + E_{D_{+}}$.
(iv) Even more interesting is that $\Delta E \simeq e \Delta \bar{V}_{z}$ in ~\ref{fig1}(b). (v) The effect of h-BN spacer layers on the electronic properties of graphene is minute. Hence, the small lattice mismatch between graphene and h-BN layers is not important for the purpose of the present study.

These features by themselves demonstrate that our model of NDC is in fact appropriate and precise. Even if the adjacent layers are coupled, the graphene layers A and B are isolated and electronically decoupled.  It should be noted that the capacitive behavior attained above by applying perpendicular electric field is equivalent to the reversed situation, where the $\pm Q$ charge stored in different plates can induce the same $\vec{E}$.

\begin{figure*}
\includegraphics[width=16cm]{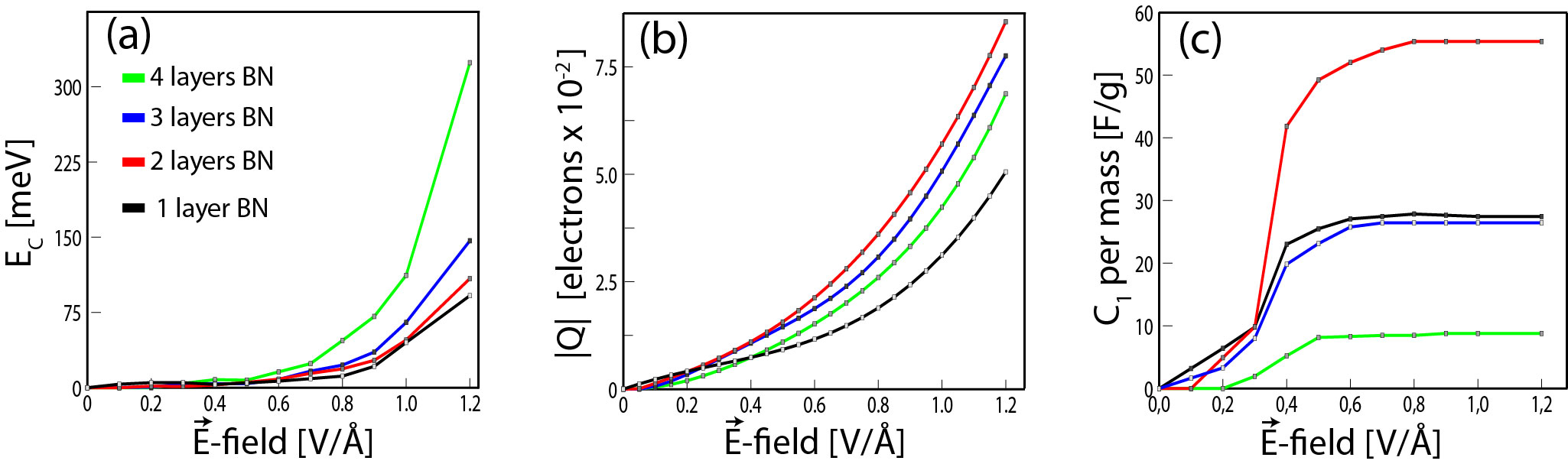}
\caption{(Color online) Variations of stored energy $E_C$, charge $|Q|$ and capacitance $C$ as a function of external electric field $\vec{E}$ calculated for different $n$ number of h-BN layers between two graphenes. Note that the capacitance values start to saturate after $\vec{E}>$0.35 V/\AA~ and reach their steady state values for $\vec{E}>$ 0.6 V/\AA.}

\label{fig3}
\end{figure*}

\begin{table*}
\caption{Number of BN layers between the graphene plates, $n$; calculated and optimized values of the distance between graphene layers capping h-BN layers, $d$ (in \AA); total mass of the primitive unitcell, $m$ (in $kg \times 10^{-22}$); calculated dielectric constants of the layered h-BN sheets, $\kappa$; magnitude of the excess charge on the graphene plates, $|Q|$ (in electrons); energy stored in the primitive unitcell, $E_C$ (in eV); local potential difference between the graphene plates, $\Delta \bar{V}_z$ (in $V$); gravimetric capacitance in Farads per grams calculated using (i) the $E_C$ and $Q$ values obtained from DFT calculations, \textit{i.e.} $C_{1} = {Q^2/2mE}$; (ii) the $\Delta \bar{V}_z$ and $Q$ values obtained from DFT calculations, \textit{i.e.} $C_{2} = {Q/m\Delta \bar{V}_z}$ and (iii) using the classical Helmoltz expression, \textit{i.e.} $C_{3} = \kappa \epsilon_{0} \frac{A}{md}$, where $\kappa$ is the dielectric constant value for bulk BN, $\epsilon_{0} = 8.85 \times 10^{-12}$ $F/m$ is the permittivity of free space and $A = 5.25 \times 10^{-20}$ $m^2$ is the area of the graphene plate in the primitive unitcell. The relevant values are given for the maximum capacitance values obtained for each $n$. The masses of the NDC models are calculated by adding the atomic masses of B, C and N  atoms in the primitive unitcell of the optimized composite systems.}
\label{table: 1}
\begin{center}
\begin{tabular}{p{0.3cm}p{0.85cm}p{0.85cm}p{0.85cm}p{0.85cm}p{0.85cm}p{0.5cm}p{0.85cm}p{0.85cm}p{0.85cm}}
\hline  \hline
$n$ & $d$ & $m$  & $\kappa$  & $|Q|$  & $E_C$ & $\bar{V_z} $ & $C_1$ & $C_2$ & $C_3$ \\
\hline
\hline
1 & 6.21  & 1.21  & 1.59 & 0.035 & 0.033 & 2.2 &  24.5 & 20.9 & 33.9 \\
2 & 9.52  & 1.63  & 2.17 & 0.060 & 0.033 & 1.8 &  54.4 & 32.8 & 16.5 \\
3 & 12.23 & 2.04  & 2.75 & 0.055 & 0.051 & 3.5 &  23.2 & 12.3 & 10.2 \\
4 & 15.14 & 2.46  & 3.29 & 0.040 & 0.093 & 4.6 &  5.6 &  5.7  & 6.9 \\
\hline
\hline
\end{tabular}
\end{center}
\end{table*}

\subsection{Capacitance}

Having tested our NDC model of capacitor, we next investigate its capacitive behavior. To this end we calculate the stored energy as a function of the applied electric field for different numbers $n$ of parallel h-BN . The energy difference between the total energies of the structures under external field and under zero field gives us the energy stored in the capacitor, $E_C (n,\vec{E}) = E_{T}(n,\vec{E})- E_{T}(n,\vec{E}=0)$. The total energies, $E_T$, are obtained by SCF energy calculations of the optimized structures. In ~\ref{fig3}(a) the variations of capacitor energy with applied $\vec{E}$ are plotted for $n=$1-4. The variation of $E_C$ is not monotonic with $n$ for the reason explained in the forthcoming part. In~\ref{fig3}(b), we present the variation of the magnitude of the excess charge, $|Q|$ stored in either graphene layers as a function of the applied field for $n=$1-4. Here $|Q|$ is also obtained from first-principles calculations by integrating the net charge in either graphene layer. In~\ref{fig3}(c) we calculate the capacitance per mass $C_1$ from the expression $C_{1}= Q^{2}/(2mE_{C})$, where $m$ is the total mass of B, N and C atoms in the model. We note that for $\vec{E} > $0.35 V/\AA, the calculated values of capacitance $C_{1}$ begin to saturate to different values depending on $n$ and become independent of the applied field as one expects. Thus, our choice of $\vec{E}=$1V/\AA~ in ~\ref{fig2} was reasonable, since it is within the saturated region for $n=$2. This very important result shows that even if the energy of NDC is obtained from the first-principles calculations, the behavior of the calculated capacitance complies with its definition. The transient behavior of $\vec{E}$ in the range of 0$< \vec{E}<$0.45  is due to the uncertainty in the calculation of optimized $E_{T}$. The gravimetric capacitance of $n=$2 is calculated to be $C_{1}=$ 54.4 F/gr, which is considered to be in the range of (EDLC) supercapacitors. The capacitance values can also be acquired from the definition, namely $C_2 = Q/m\Delta \bar{V}_z$ using the calculated charge values $|Q|$ in ~\ref{fig3}(b) and $\Delta \bar{V}_z$ from the plane-averaged potential $\bar{V}(z)$ as described in ~\ref{fig1}. The capacitance values acquired this way are in the range of those calculated from $C_{1}= Q^{2}/(2mE_{T})$ in ~\ref{fig3}(c). Nonetheless, in the rest of discussions we used the capacitance values $C_1$ obtained from the quantum mechanical calculations of energy and charge, since due to the plane averaging process of electronic potential, $\Delta \bar{V}_z$ does not provide the necessary precision compatible with that of the total energy calculations. In Table I, we list various parameters and calculated values relevant for the calculations of gravimetric capacitances per unit mass. We do not consider the graphene bilayer corresponding to $n=0$, since the charge stored in the layers are shorted.

\begin{figure}
\includegraphics[width=8cm]{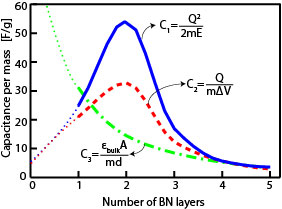}
\caption{(Color online) Comparison of the calculated capacitance values of NDC at $\vec{E} =$ 1.0 V/ \AA~ as a function of the number of insulating h-BN layers, $n$. Capacitance values calculated using the optimized total energy $E_{T}[n,\vec{E}]$ obtained from DFT, $C_1$; using the plane-averaged $\Delta \bar{V}_z$, $C_2$; and the purely classical Helmholtz formula, $C_3$; are shown by  solid(blue), dashed(red) and dash-dotted(green) lines, respectively. The dotted lines show the hypothetical capacitance values for $n<1$.}
\label{fig4}
\end{figure}

In ~\ref{fig4} we compare capacitance values calculated using DFT and classical Helmholtz model as a function $n$. The capacitance value acquired from the DFT results is small for single h-BN layer; but passes through a maximum and then decreases with increasing numbers of h-BN layers. The latter behavior for large number of h-BN layers is reminiscent of that of classical capacitance, namely $C \propto 1/d$, since the distance $d$ between the charged capacitor plates also increases as $n$ increases. However, that the increase in the value of $C_1$ as $n$ increases from 1 to 2 is surprising, since the effect of the insulating medium consisting of h-BN layers is indigenous to the DFT calculation of $C_1$ deduced from $E_{C}$. This situation is attributed to a quantum size effect as revealed from our analysis of high frequency dielectric constant, $\kappa_{\nu=\infty}$, as well as low frequency (static) dielectric constant $\kappa_{\nu=0}$, of h-BN sheets calculated as a function of $n$. Because of the layered character, one distinguishes the in-plane ($\parallel$) and perpendicular ($\perp$) dielectric constants in BN spacer. The calculated $\kappa_{\nu=0,\parallel}$ and $\kappa_{\nu=0,\perp}$ of the layered bulk BN crystal are 5.25 and 3.80, respectively, which are in fair agreement with experiment.\cite{expdi} Calculated values of $\kappa_{\nu=0,\parallel}$ ($\kappa_{\nu=0,\perp}$) are listed in Table I for $n=$1-4. Accordingly, $\kappa$ increases with increasing $n$ and hence displays a strong size effect. Rather small values of $\kappa$ for $n=$1 and 2 cause that $C_1$ first increases with increasing h-BN layer for $n \leq $2, but it decreases with increasing $n$ for $n > $2 as shown in~\ref{fig4}. In other words, although a decrease in the capacitance value is expected as the layer-layer separation increases, the increase in the dielectric constant of h-BN layers counters this effect. Hence, as a result of the competition between the increased layer-layer distance and increasing dielectric constant values, the capacitance value passes from a maximum at an optimum spacing after which it starts to decrease again. In the same figure we also plotted the variation of capacitance obtained from the Helmholtz model $C_{3}=\kappa \epsilon_{0} A / md$, where $d$ is the distance between graphene layers corresponding to a given $n$, and $\kappa$ is the dielectric constant of bulk BN. This is a purely classical value. We see that quantum effects dominate the capacitance for $n < $4, while they decay quickly and the capacitor behaves as a classical capacitor for $n \geq $4.

\section{Conclusions}

We emphasize the important aspects of the present study by way of conclusion: (i) We proposed a model of a nanoscale dielectric capacitor (NDC) composed of two metallic graphene layers separated by an insulating medium comprising a few h-BN layers. Graphene as well as h-BN layers are in registry. BN was chosen as a dielectric because it is an excellent spacer with a lattice constant close to that of graphene. Graphene and h-BN are two well-known single layer honeycomb structures; either one of them can grow easily on the other. Therefore, the proposed model can easily be fabricated. (ii) We showed that our model can attain high capacitance values. This is crucial for high capacity energy storage. Even in classical region for $n \geq$ 4 the capacitance values $C_3$ are still very high due to the separation of graphene layers at nanoscale. (iii) The DFT method used to separate charges of different polarities between two metallic plates and to calculate the optimized total energy is original and allows us to treat the nanoscale dielectric capacitor from the first principles. (iv) The analysis on the spurious vacuum charging is crucial for the calculations from the first-principles using periodic boundary conditions, as it is in the present study. These spurious effects have been eliminated by using local basis sets. (v) We also showed that quantum effects become crucial at nanoscale and how they recede as the dimensions of the capacitor increase. Thus, the present model of NDC keeps the promises of future applications. Our model of nanocapacitor allows the sequential and multiple combination of graphene/BN/graphene/BN/...../graphene to achieve high volumetric densities and to attain high capacity energy storage.  Apart from the graphene/h-BN/graphene composite nanocapacitor, layered composite materials composed of graphene like insulators GaAs, AlAs, InN\cite{ansiklopedi} etc. which are capped by  single layer metallic Si and Ge (i.e. silicene and Germanene)\cite{silicene}, or layered transition metal dichalcogenides\cite{mx2} can be used for the same purposes. Finally, we point out another interesting situation, where the metallic plates and the dielectric medium between them can be arranged on the same single layer honeycomb structure consisting graphene and h-BN domains. This way a 2D capacitor can be realized and can be integrated to the nanocircuit on the same layer. The recent studies achieving the growth of graphene and h-BN continuously on the same layer\cite{sutter2012} indicate that this conjecture can be promising in near future. We hope that the present study will initiate further experimental and theoretical studies in this field.

\section{Acknowledgements}
The authors thank Can Ataca for his assistance in the calculation of the dielectric constants of h-BN layers. The computational resources have been provided by TUBITAK ULAKBIM, High Performance and Grid Computing Center (TR-Grid e-Infrastructure) and UYBHM at Istanbul Technical University through Grant No. 2-024-2007. This work was supported by TUBITAK and the Academy of Sciences of Turkey(TUBA).

\bibliography{capacitor_arxiv.bbl}

\newpage

\begin{figure*}
\includegraphics[width=16cm]{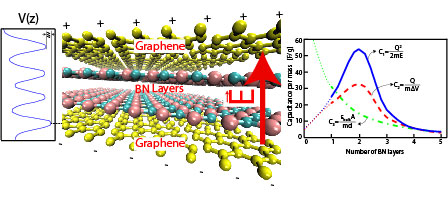}
\begin{center}
\caption{(Color online) TOC Figure}
\end{center}
\label{TOC}
\end{figure*}

\end{document}